# A Method for Estimating the Probability of Extremely Rare Accidents in Complex Systems

Ítalo Romani de Oliveira and Jeffery Musiak

**Boeing Research & Technology**

*Abstract* – Estimating the probability of failures or accidents with aerospace systems is often necessary when new concepts or designs are introduced, as it is being done for Autonomous Aircraft. If the design is safe, as it is supposed to be, accident cases are hard to find. Such analysis needs some variance reduction technique and several algorithms exist for that, however specific model features may cause difficulties in practice, such as the case of system models where independent agents have to autonomously accomplish missions within finite time, and likely with the presence of human agents. For handling these scenarios, this paper presents a novel estimation approach, based on the combination of the well-established variation reduction technique of Interacting Particles System (IPS) with the long-standing optimization algorithm denominated DIviding RECTangles (DIRECT). When combined, these two techniques yield statistically significant results for extremely low probabilities. In addition, this novel approach allows the identification of intermediate events and simplifies the evaluation of sensitivity of the estimated probabilities to certain system parameters.

## I. Motivation and Structure of this Work

Heterogeneous organizations having both technical infrastructure (hardware and software), as well as human beings filling certain roles, constitute complex systems effecting our daily lives, such as air traffic management systems, power generation systems and even financial systems. Despite increasing levels of automation, there are still human individuals responsible for their operations, who are influenced by their internal states and processes and by external inputs. A term 'socio-technical system' is sometimes used as the individuals interact with the technical subsystems as well as with other individuals participating in the large system,

Multi-agent dynamic risk models (MA-DRMs) have been proven successful for analyzing complex socio-technical systems in regard to safety properties [1], [2]. The power of such models becomes clear when used in conjunction with simulation methods in order to estimate failure and accident rates. These rates have to be estimated during early design stages, particularly for the systems whose failure could result in catastrophic consequences. Furthermore, these rates need to be under certain target levels of safety (TLS). This requirement is established by regulations such as [3]–[5], and is even more important as the systems become more autonomous. One of the difficulties in this estimation is that the TLS, expressed as a probability of occurrence of an undesired event per unit of time, is often on the order of $10^{-9}$ or below, and hardly can be checked analytically with realistic models such as those considered in this paper.

One way to proceed with this estimation is to use sequential Monte Carlo methods [6], [7]. One of the best known sequential Monte Carlo methods is referred to as "multilevel" or "importance splitting" [8]. This family of methods requires the system model to be a Markov process, which can be achieved in many practical cases [9], [10]. Besides that, in order to become computationally efficient, sequential Monte Carlo solutions take the form of particle filtering techniques, as is the case of the Interactive Particle System (IPS) [11], [12]. However, developing and tuning a proper particle filter to deal with rare events is challenging because of the degeneracy that results from the lack of diversity after successive re-sampling steps, which in turn results in high variance or, not rarely, in not obtaining any occurrence of the target events.

Another way to perform rare event sampling is to employ optimization techniques such as the Cross-Entropy method [13], [14], in order to find alternative probability distributions, with higher occurrence of the desired event, which then can be used in importance sampling [15] to allow the estimation of the base rate of occurrence of the rare event. However, two important drawbacks of these schemes must be taken into account: first that many optimization algorithms, including the Cross-Entropy, are not guaranteed to find the globally optimum points; second, even if an optimization algorithm is guaranteed to converge to a globally optimum point, such algorithms usually converge to very small regions of the search space and thus other significant regions may be neglected. In rare event probability estimation, these features may result in errors of many orders of magnitude.

The family of methods named Markov-Chain Monte Carlo (MCMC) [16] has achieved a prominent success in failure analysis of complex systems, having being used in various ways. In its more flexible type of use, it is used to estimate parameters of an importance sampling (IS) distribution in high dimensional problems, providing good approximations of the rare event probability distribution with a low number of function evaluations [17]. This method requires starting with a point in the failure region, which usually can be obtained with engineering judgment, and using a generic transition probability density function to generate samples from an evolutionary Markov Chain which converges after a number of samples. The main advantage of this approach is to mitigate the influence of the dimensionality (i.e., the number of input random variables) in the number of sample instances needed, although not eliminating it. In principle, this approach is convenient to use for one of the case studies of this paper, where the number of random variables is very high; however, because these variables govern the behavior of a feedback aircraft control loop, the aircraft may become undesirably unstable during the MCMC convergence, so we took a different approach.

In its more specific type of use [18], the MCMC is used within the evaluation of the limit state function, with the advantage of modelling intermediate events and enabling optimization of the sampling via subset simulation. However, such approach requires the system under analysis to have states with stationary probability distribution, which is hardly true for socio-technical systems that depend on human agents and do not run on a continuous basis. For example, when studying transportation systems from the perspective of independent vehicles, it is natural to model the system operation as starting at the beginning of a trip (or vehicle mission) or set of trips, or some finite time before them, and finishing at the end of the trips, hence a finite-time operation. This feature does not fit well to the principle of convergence after many iterations used by the MCMC in the limit state function.

Another candidate method for tackling our estimation problem comes from the field of Structural Safety, and is called the "extrapolation method" [19], [20], in which a reliability index is computed with extremized input values at lower computational cost, and then extrapolated to obtain the reliability index corresponding to input values with unitary scale, based on asymptotic properties. The sampling needed can be highly optimized by using multiple models as in [21] and this approach also copes well with high dimensionality. Yet, a limit state function with complex shape and very high reliability indices can pose challenges to it.

Still other methods have been developed for rare event probability estimation, and a good summary on them can be found in [17]. Our novel approach emerged naturally from the applications we have been studying, involving MA-DRM models.

The difficulty of performing timely computations for rare event probability estimation may be one of the main explanations for the scarcity of the use of MA-DRM models for risk assessment of new system designs. However, it is advocated here that this should not be a reason to discard the power of such methods. This paper demonstrates that there is a practical way to estimate very rare event probabilities with MA-DRM models of complex systems, if one chooses a model with a complexity level matching the available computational power. The basic principles of this approach were published in [22], and here an in-depth presentation developed.

After this introductory section, the contents of this paper is organized as follows: Section II presents a basic version of the probability estimation method; Section III presents a case study on a hypothetical aircraft operation, providing concrete explanations on the basic method; Section IV introduces the more general estimation method, allowing Stochastic Differential Equations (SDEs); Section V, analogous to Section III, explains the method in more concrete terms with an expanded version of the same hypothetical example; and Section VI contains the conclusions. There are appendices with more details on validation and comparisons of our algorithms, and a section defining the mathematical symbols in the end of the paper, before the list of bibliographical references.

## II. THE BASIC METHOD

The basic method for a rare event probability estimation proposed in this paper is summarized in Figure 1.

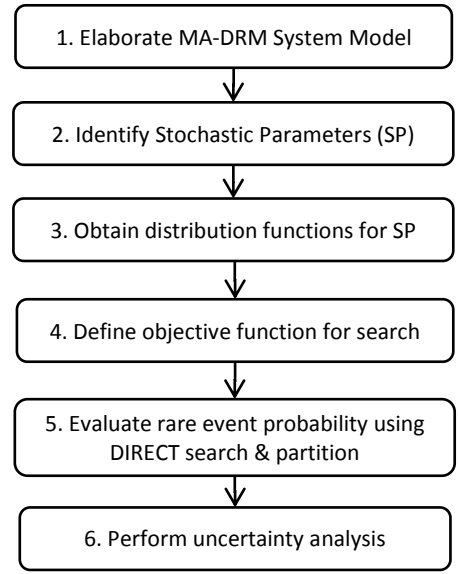

**Figure 1**: The basic method for a rare event probability estimation.

The estimation process starts in Step 1 when the system agents become defined, as well as their dynamics and interactions, based on expert knowledge on technical systems and organizations [23]–[25]. Besides the situational awareness of the agents, this model represents the occurrence of faults in technical components and environmental dynamic conditions, which may lead to systemic hazards and generate failure or accident events. Often, it is convenient to start with a pre-existing knowledge of hazards in analogous systems, then use that knowledge as a basis for the definition of agents and the breakdown of main components [26]. This approach mitigates the negative effect of simplifying assumptions, which are inherent to any model elaboration process and to the models themselves. It also minimizes the loss of unstructured, intuitive and experience-acquired forms of knowledge, which might escape from a purely structural approach.

As this work concerns estimation of rare events arising from the MA-DRM, the model has to be mathematically sound, and the type of mathematical model to be applied depends on the complexity of the system to be modelled. For example, a very elementary system with only one agent (e.g. a bouncing ball) can be modelled as Ordinary Differential Equations (ODE) with continuous variables. However, the class of systems aimed here has at least one discrete variable (a mode variable), so it is fair to say that at least some sort of Hybrid Automaton [27] or colored Petri Net [28] has to be used. Also, if the system has multiple agents, some composition and inter-communication representation must be available in the formalism [29]–[31]. As the system complexity increases, the system elements may be governed by sets of SDEs that can be activated or deactivated at any given time. The process of activating and deactivating equations is called switching and, when switching occurs by hitting a boundary or by stochastic jump processes, the system can be mathematically considered a General Stochastic Hybrid System [32]. The execution of such system is a General Stochastic Hybrid Process (GSHP), and can be conveniently modelled by Stochastically and Dynamically Colored Petri Nets (SDCPN) [33]. The method presented in this section, however, needs the assumptions that the model has no SDEs or, in other words, has no embedded Brownian motion. The full complexity of GSHPs will be tackled by an enhanced version of the method, to be presented in Section III.

Step 2 of Figure 1 consists of identifying the set of purely stochastic variables of the model, where "purely" means that each variable must not have in their definitions operations involving other model variables nor a previous value of itself (that is, they must not have a recursive definition). Because of this feature, these variables can be considered as input variables and can be called system parameters or model inputs. These parameters must follow some known probability distributions and this is the reason for performing Step 3, which determines their probability distribution functions. This determination may be based on expert knowledge or data analytics. If the stochastic parameter is discrete, it has to be represented as a continuous variable in order to proceed with the next steps of this method.

In Step 4, a scalar-valued objective function with multiple inputs is chosen and becomes a central element of the search algorithm, which may seek a minimum or a maximum value of the function. By way of example in a nuclear power plant, the reactor core temperature can be the objective function which leads to a catastrophic rare event when rising beyond a certain threshold. Although, the temperature itself may not be the only criterion for the catastrophic event, it is a strong indicator. In the case of collision between aircraft, a miss distance (or distance at the closest point of approach) can be an objective function (min-valued) as examples in this paper.

Having defined the objective function, the method proceeds with Step 5, where an algorithm is used to search for the minimum (or maximum) values of the objective function and, at the same time, to partition the search space in order to facilitate probability calculation. The DIviding RECTangles (DIRECT) [34] algorithm used in this study provides a partition of the search space that, if sufficiently refined, can be used in combination with the probability distribution function of the search variables (either density or cumulative distribution function), in order to obtain an accurate estimation of the probability of occurrence of the event being sought. This approach differs from Monte Carlo because it uses no sampling for the stochastic parameters.

Finally, Step 6 examines the uncertainty in the knowledge of the moments of the probability distribution functions of the stochastic parameters. This uncertainty raises mainly two questions: i) What is the confidence level of meeting the TLS? And ii) which stochastic parameters of the rare event are most sensitive, and with how much intensity? The answer to the first question determines whether or not the current system design concept is acceptable, and the answer to the second question helps finding ways to improve the current design concept and is performed by means of sensitivity analysis techniques. Standard statistics has plenty of methods for providing these answers but, in one way or another, these methods require the re-calculation of the rare event probability with different inputs. This goal is greatly facilitated by the partition of the parameter space provided in Step 5, diminishing the necessity of re-executing simulations of the system model.

Section III illustrates a case study of the basic method introduced in this section.

### III. CASE STUDY USING THE BASIC METHOD

The estimation approach from Section II is aimed at hybrid systems with some complexity. One case of such systems was chosen to be used for demonstration of this approach, presented in the next sub-sections.

*Step 1: Elaborate MA-DRM model*

This application case consists in the operation of a transport aircraft in a certain phase of flight, described at high level by a multi-agent system, illustrated in Figure 2.

**3**

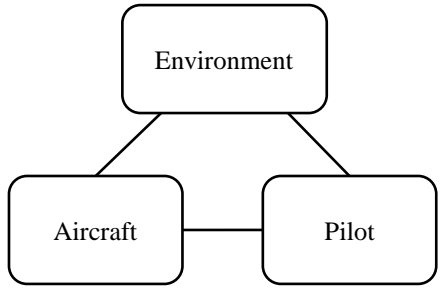

**Figure 2**: Multi-Agent aircraft operation model.

The Aircraft agent is modelled as a point-mass aircraft described by an ODE system of six equations in a feedback control loop, subject to independent inputs from the environment and from the pilot. This part of the model is based on [35] and the details of which are omitted because they are not relevant to the main contribution of this paper.

The control inputs are determined by a hybrid feedback controller, which can be either in automatic mode or subject to pilot inputs. The automatic mode is programmable by means of a sequence of 3-D waypoints and segment modes, and will keep the aircraft on the desired path even in the presence of wind; there might be several pilot inputs and input modes to control the aircraft, however, the only pilot intervention allowed is to command an emergency maneuver of full-thrust climb in this example. The parameter values used to fill the model of aircraft dynamics in this case study correspond to a commercial single aisle jet aircraft.

The programmed flight path for the experiments herein is illustrated in Figure 3. The aircraft enters the scenario at the upper right corner and follows a predefined route with a "U-shape," which descends and passes between two peaks in the terrain. The aircraft flies until either accidentally hitting terrain, going out of the airspace bounding box, or reaching a maximum flight time $T$, whichever is the first to occur. The idea of such route is to resemble a flight approaching an airport in the proximity of mountains. In nominal conditions, the trajectory terminates at the lowermost waypoint, from which the aircraft will proceed to the final approach. In a risk assessment, which is recommended to happen before the use of a route with similar features in real life, there is interest in the non-nominal conditions, which are triggered by errors or faults that, in extreme cases, lead to an accident, which in this example is hitting terrain. In order to hit terrain, the minimum distance between the terrain and aircraft, $d_{\min}$, has to be 0. If the operation goes correctly, without wind and other disturbances, the outcome for $d_{\min}$ is 1,354 ft., determined by model simulation.

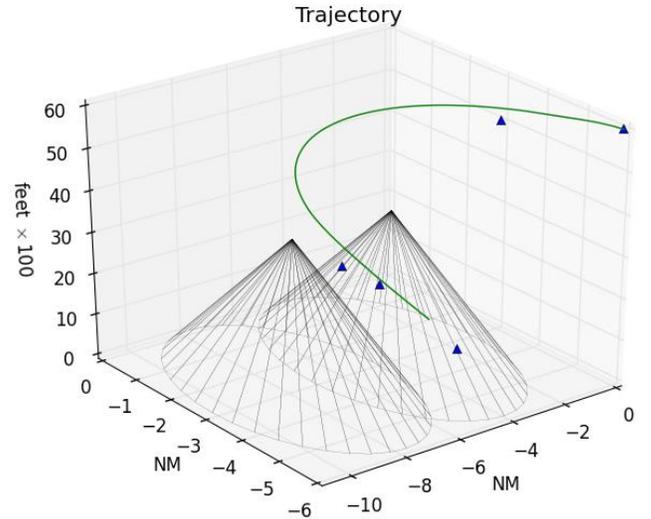

**Figure 3**: Illustration of the aircraft operation model.

For the purpose of risk assessment, it is assumed that the aircraft has an altimetry fault when it enters in the scenario, which is manifested in a numeric error $\epsilon_a$, which influences the trajectory flown.

The *Environment* agent is composed of terrain and atmosphere. In the example model, the terrain has a base of altitude 0, on which lay two cones with base radius of 3 nautical miles and height of 3600 feet, positioned as shown in the figure. If the distance between terrain and aircraft reaches the minimum value $m = 0$, the simulation is immediately stopped. The atmosphere acts with a constant wind with horizontal x-y velocity components $w_x$ and $w_y$.

The *Pilot* agent's role is only to detect the altimetry fault and, in response, initiate the avoidance maneuver. As a cognitive agent, he or she is capable of detecting the fault either by his/her interaction with the Environment (e.g., look out of the window and observe terrain), or by interaction with the aircraft itself, checking other instruments such as radio altimeter, computer messages, etc. The time that the pilot takes to detect and react to erroneous altitude is stochastic and denoted by $t_r$.

Aircraft altimetry systems are nowadays very reliable and further are protected by having redundant systems. Still, faults happen in some cases, including icing or other types of sensor obstruction, computing error, etc. Whatever the phenomenon is, it may happen and, in the present model, it is established that, when it happens, the altimetry system will present the above mentioned altitude error $\epsilon_a$, which in turn causes the aircraft to fly with an altitude offset $\epsilon_h$, of same magnitude and opposite value, i.e., $\epsilon_h = -\epsilon_a$. The probability that this fault actually occurs is beyond the scope of this model and is left to more concrete research. The model just assumes that the aircraft enters the scenario with this altitude reading error and that the flight guidance system makes the aircraft fly with the altitude shifted by this error value, until the flight crew detects the altimetry fault by some means. It is then that the pilot exercises a contingency maneuver to make the aircraft climb steadily at maximum thrust, for a certain time, in order to be clear of

**4**

terrain. This contingency maneuver is simplified, but ensures that the aircraft will climb at its maximum performance, without stalling, until it is above the altitude of the peaks. This is illustrated in Figure 4, in two different perspectives. The initial continuous part of the trajectory is the navigation with erroneous altitude, and the dashed part is the avoidance maneuver.

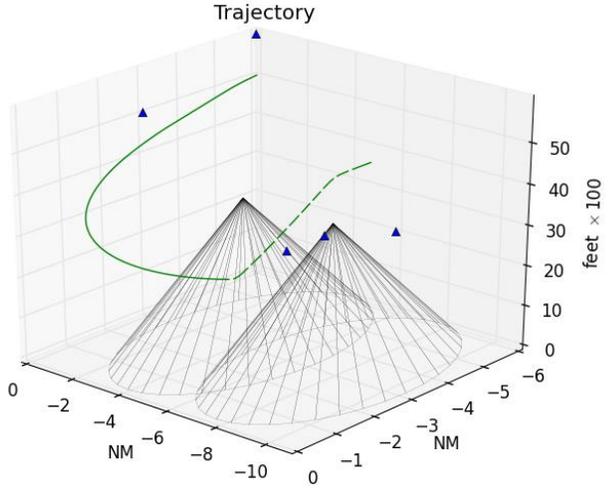

(a) Perspective I

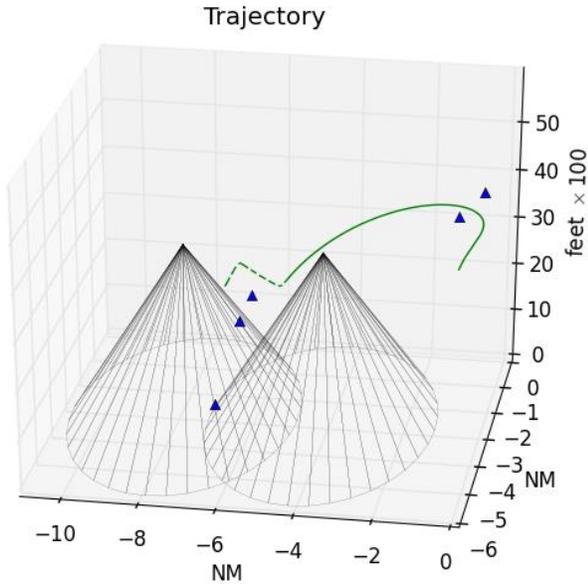

(b) Perspective II

**Figure 4**: Avoidance maneuver after altimetry fault, in two different perspectives.

This completes Step 1. The next step of the estimation approach is to use some of the variables of the system model to perform a search which will lead to the occurrence of the events of interest. The method for performing this search is described in the following section.

*Step 2: Identify Stochastic Parameters (SP)*

The altitude offset $\epsilon_h$, caused by an altimetry fault, and the time $t_r$ for crew reaction, are identified as the stochastic parameters from the model above.

*Step 3: Obtain distribution functions for the stochastic parameters*

The determination of the probability distribution function of the stochastic parameters should be done by data analytics or by experts in a guided process, such as one described in [36], which has systematic reduction of bias and inconsistency. However, as this case study is hypothetical, the determination here is for illustrative purposes. The time that the pilot takes to react to the altitude error, $t_r$, is associated to an exponential distribution with the mean $\mu$ of 30 seconds; the altitude offset $\epsilon_h$, the other stochastic parameter, is associated to a normal distribution with moments $\mu = 0$ and $\sigma = 100$ feet. The constant wind velocity components $w_x$ and $w_y$ are normally distributed with moments $\mu = 0$ and $\sigma = 15$ knots.

*Step 4: Define objective function for search*

The objective function defined here will be used by the search and partition algorithm in the next step of the method. When searching for the regions of the parameter space where collision with terrain happens, a natural choice for objective function is the terrain miss distance in the example model, denoted by $d(x)$, where $x$ is the vector of parameter values. "Miss Distance" means the minimum distance that ever existed between aircraft and terrain in one scenario execution. If $d(x) = 0$, it means that the aircraft collided, so the target event happened. The algorithm then uses this distance to find successive combinations of parameter values to determine where the collision actually happens, i.e., parameter values in the so-called target region. Each evaluation of $d(x)$ requires running the model because it is not possible to obtain it analytically. This function is also referred to as limit state function in analogous works, such as [19].

The algorithm best suited for searching and partitioning the parameter space for the purposes of this method is called DIviding RECTangles (DIRECT) [34], with some modifications explained later in this paper. One of the most important modifications is to adapt the objective function which was taken naturally from the system, adding some extra rules in order to reduce errors in the computation of the net probability of the target event.

After performing the assigned number of interactions, DIRECT will partition the search space in a number of rectangles (hyperboxes of two dimensions) corresponding to the number of objective function evaluations. In order to demonstrate this partition in a figure, it is assumed that the wind velocity be zero and only the variation of $(t_r, \epsilon_h)$ is considered. Figure 5 shows 100 rectangles generated by the algorithm, superimposed to the contour plot of the basic objective function, the terrain miss distance, where on the lower right corner it is possible to see the combination of parameter values that end up in collision with the ground.

Each evaluation of the objective function in the DIRECT algorithm corresponds to a rectangle's centroid, but the shaded contour plot was made independently of DIRECT, requiring a separate set of one thousand function evaluations (the plot function smoothed the colors and the individual evaluations cannot be perceived).

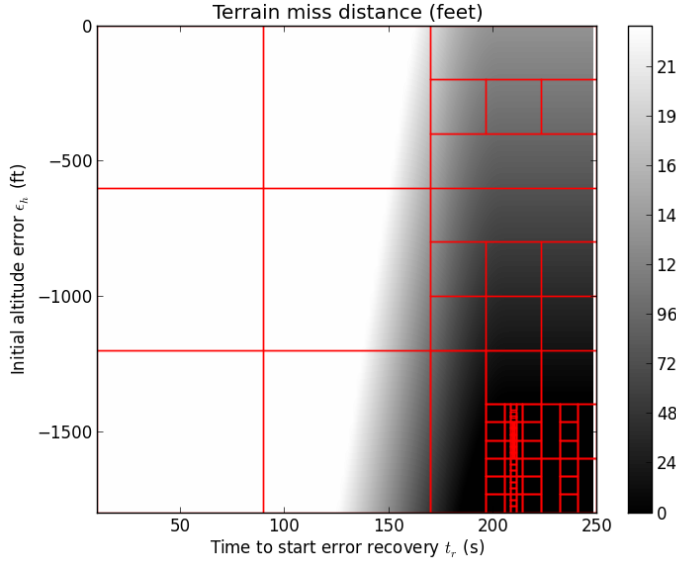

**Figure 5**: Contour plot of the terrain miss distance function $d(x)$, on the background, and search rectangles generated by the DIRECT algorithm.

It can be seen that the algorithm makes the partition finer in the regions of smaller objective function values. If more objective function evaluations are performed, the rectangles become smaller and smaller. As it can be noticed in the figure, the rectangle subdivisions are concentrating in a small spot of the target region ($d(x) = 0$). However, the target region actually spans over a larger area, roughly defined by $\{t_r \geq 200 \land \epsilon_h \leq -1400\}$, and it is wasteful having heterogeneous distribution of rectangles in this area of constant objective function. If DIRECT reaches such a basin, it has no proper criterion to guide subdivisions in it, thus this undesirable concentration happens. Instead of appending new logic to the DIRECT algorithm, it is possible to improve rectangle subdivision by hybridizing the objective function.

Usually, the probability of the rare event is not uniform in the target region basins, and it would be good that the rectangle subdivision be finer in the regions of higher probability, in order to decrease discretization error. Thus, a new objective function $f_o$ can be defined as:

$$f_o(x) \triangleq \begin{cases} d(x) & \text{if } d(x) > m \\ -g(x) & \text{if } d(x) \leq m \end{cases} \quad (1)$$

where m is the distance threshold which defines the rare event (here m = 0), and $g(x)$ is the probability density function of the stochastic parameters. The first case of the function guides the search outside the target region, while the second case does it inside, converging to the places with higher probability density. A plot of the function $g$ is given in Figure 6, for a null wind.

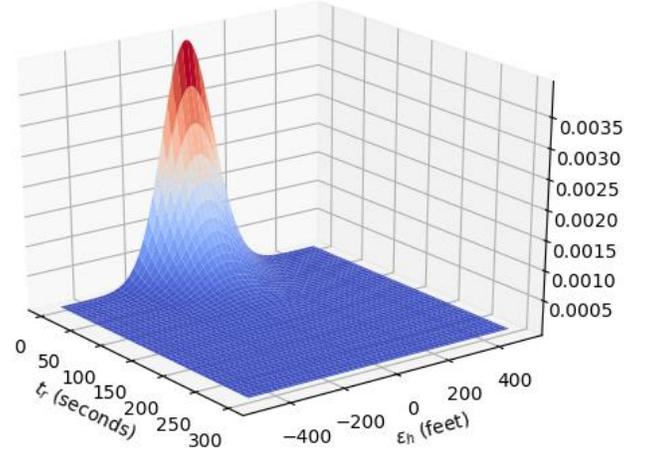

**Figure 6**: Probability density function $g(t_r, \epsilon_h | w_{x,y} = 0)$.

Once a point is found inside the target region, the subsequent divisions are led to concentrate at the highest regions of $g$, and this works well under the condition that $g$ is both continuous and devoid of flat basins inside the target region. With these conditions holding, convergence of DIRECT stays guaranteed, because $g$ has a maximum and thus $f_o = -g$ converges to its minimum in a closed interval, provided that the target region is continuous around the argument of this minimum. An example of this convergence happening is shown in Figure 7.

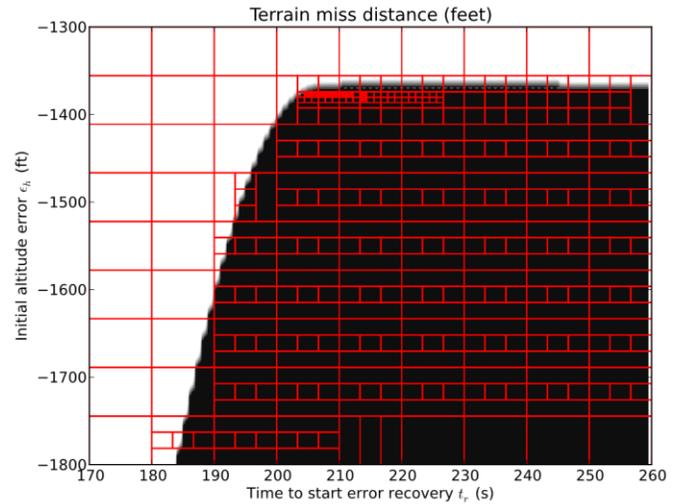

**Figure 7**: Convergence of the DIRECT algorithm with inner-guided objective function, with $m = 0$.

The contour color scale on the background was changed to make the target region, in black, more distinct from the remainder of the points, in white. One can note a concentration of rectangles with high probability density at the upper left corner of the target region. In addition, it is possible to see that, at the opposite corner, there are bigger rectangles. This difference more or less shows the gradient of the density function $g$.

If a more conservative probability estimate is needed, it is possible to increase the target region threshold $m$, so that the rectangular division is concentrated on the outer edge. Using $m = 25$ ft, instead of zero, the rectangles that were obtained are shown in Figure 8. The grey area delimits the expansion of the target region, between 0 and 25 ft.

This area of concentration occurs because the probability density on the tails of the probability distribution functions decays exponentially for $t_r$, and even more strongly for $\epsilon_h$, hence the probability mass is concentrated towards the central moments of the distribution functions (the wind variables do not influence this distribution because the wind vector has to be fixed in order to obtain figures 5-8).

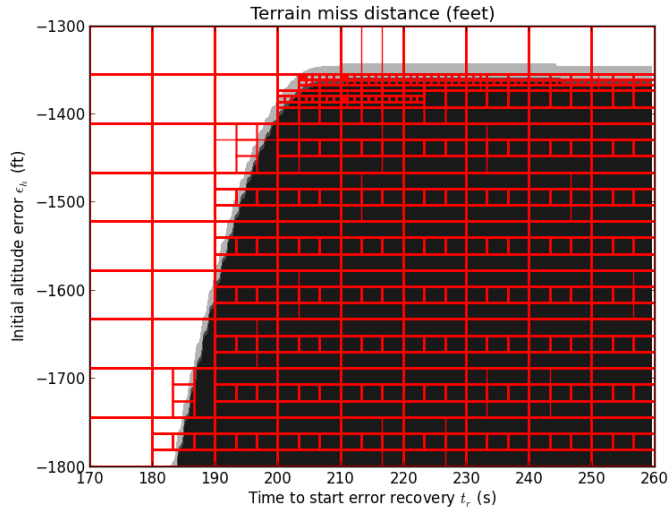

**Figure 8:** Convergence of the DIRECT algorithm with inner objective function, with $m = 25$.

*Step 5: Evaluate rare event probability using DIRECT search & partition*

The elements presented in the previous sections are used here to perform the evaluation of the rare event probability. As the number of dimensions and the complexity of the system model strongly affects the performance of the search & partition algorithm, two computation sessions are presented here to illustrate scalability issues. The first session uses only two stochastic parameters, and the second session uses all four stochastic parameters defined above.

The stopping criterion of the DIRECT search & partition algorithm was arbitrated as completing a number $q$ of evaluations during which the change in the estimated probability $p_k$ is no more than a fraction $\varepsilon_M$ of this probability, i.e., $|p_{k+j} - p_k| < \varepsilon_M . p_k, j = 1, \ldots, q$, with values $q = 10^3$ and $\varepsilon_M = 10^{-9}$. The implementation of the DIRECT algorithm is based on the NLOPT library (this acronym stands for Non-Linear Optimization) [37].

The first computation session allowed variation of only the parameters $(t_r, \epsilon_h)$, with the wind parameters fixed at zero. Executing the DIRECT search & partition algorithm and registering partial results on the probability of the target event (collision with the ground), the evolution of their estimated probabilities along the algorithm iterations is shown in Figure 9.

For this computation, the objective function $f_0$ was set with $m = 25$, according to the definitions of the previous section. The convergence stop criterion caused the computation to stop after 3,600 evaluations of $f_0$, having computed the probability of 3.24E–46 per operation.

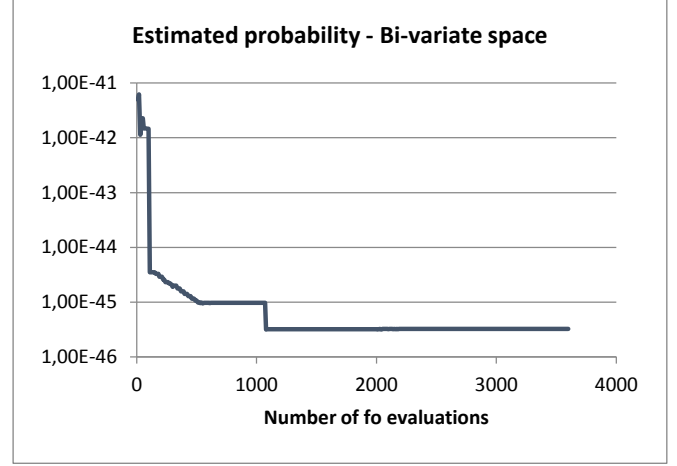

**Figure 9**: Convergence of the probability computation with DIRECT in a bivariate search space.

Now, the computation task is scaled up. The constant wind element is varied, making $d$ a function of $x = (\epsilon_h, t_r, w_x, w_y)$. The effect of such change on the probability estimation algorithm is that the number of hyperboxes required to cover the search space grows by the power of the number of dimensions $n$, therefore many more rectangles are needed to provide a partition with divisions small enough in each dimension (a hyperbox is a rectangle of dimension $n$). To thoroughly follow this reasoning, the number of function evaluations would have to be squared when going from 2 to 4 dimensions, therefore the next convergence experiment used $q = 10^6$ hyperboxes, one thousand times of what was considered enough for convergence with two varying parameters. When such high number of evaluations was run, however, it was noted that a high proportion of the generated hyperboxes were of negligible probability weight in comparison with the ones with the heaviest weights, hence unnecessary. In order to avoid these wasteful computations, another rule was added to selection of hyperboxes in the algorithm: if the hyperbox selected by the DIRECT criteria has probability lower than a fraction $\beta$ of the highest probability of a single hyperbox at the moment, this hyperbox is skipped and not divided (as the number of hyperboxes is finite, there will always be an adequate $\beta$). The algorithm will proceed to the next hyperbox according to the criteria of DIRECT and use the same rule. If it happens that all remaining hyperboxes are skipped, this algorithm will use the default option inherited from the NLOPT implementation, which is to point to the hyperboxes with largest size and, among these, choose the one with smallest objective function value.

When the algorithm was run with these settings, the convergence needed far fewer evaluations than the assigned $q = 10^6$. The results can be seen in Figure 10. The final probability value is approximately the same for the subsequent 750,000 evaluations actually performed ($10^6$ in total), dispensing with the need to show all the range of evaluations.

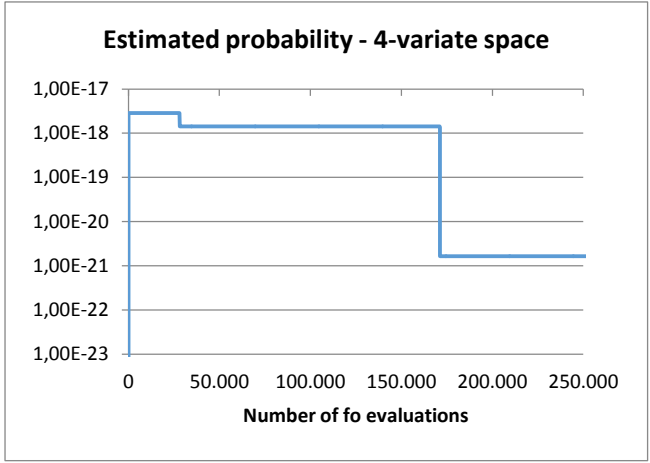

**Figure 10**: Probability values obtained with DIRECT in a 4-variate search space.

The actual computing time for these estimations is dominated by the time taken to compute $f_o$ which, in an Intel Xeon X5650, clocked at 2.67 GHz, takes on average 2 milliseconds, for a simulation program developed in C++ and compiled with GCC/Mingw. The overhead caused by DIRECT is negligible in relation to this time.

*Step 6: Perform uncertainty analysis*

The moments (e.g. mean, standard deviation) of the probability distributions of the stochastic parameters are subject to uncertainties. When a Mean Time To Fail (MTTF) for a piece of equipment is declared by the vendor, this MTTF is based on a finite number of samples or may have been calculated with compound uncertainties. In the case of the hypothetical example under study, the determination of the mean and standard deviation of $\epsilon_h$, and the mean of $t_r$ may have been determined from expert knowledge, which also embodies uncertainties. Usually, the uncertainty of each parameter is summarized as a confidence interval, the worst and best cases of the target event probability occurring at some combination of extremes of those intervals. Re-evaluating the target event probability at these points requires extra computational effort, which may be considerably reduced if the search space is already partitioned.

Instead of fully re-running Step 5 above, the partition of the search space obtained in that step can be re-used, and only the probabilities of the corresponding hyperboxes are re-calculated. This avoids the re-calculation of the objective function, which is the most expensive part of the overall computation. This is a general feature of this method. Such simplified re-calculation may cause some additional error depending on the shape of the distribution functions, if the existing partition is not a good match for the new shape, however this would be rare, as the target region usually occupies monotonic regions on the tails of the distribution functions. Finally, by making this calculation faster, the determination of the confidence intervals of the target event probability become proportionally faster, as well as the determination of parameter sensitivities.

Parameter sensitivity analysis produces valuable design information, in the sense that it is possible to identify the stochastic parameters to which the probability of the rare event in the critical system modelled is most sensitive, with the sensitivity rates calculated during this type of analysis [38]–[40]. Therefore, the system design can be changed to decrease the sensitivity of the most sensitive parameters and become more robust and less subject to uncertainties.

IV. THE METHOD FOR MODELS WITH STOCHASTIC DIFFERENTIAL EQUATIONS

The approach presented so far works well if the distribution of each stochastic parameter is known and it is feasible to obtain a partition of the search space with small enough hyperboxes, in order to minimize errors caused by the discretization of the target region. However, it is important to have in mind that complex socio-technical system have large numbers of disturbance variables and, more than that, they can be governed by discretely-switching stochastic differential equations (SDEs), for which sampling in the trajectory space is needed. This is much harder than doing so in the space of an input vector of finite size, because an SDE has infinitely many implicit random variables. As previously mentioned, such hybrid systems can be modelled as General Stochastic Hybrid System (GSHS) [32], but the basic method of Section II is not powerful enough to estimate probabilities with such models. Therefore, the estimation method for rare event probability is enhanced in this section. The overall method flow becomes as in Figure 11.

Steps 1 through 3 and 8 are identical to the basic method from Figure 1. The remaining steps have significant differences, so they are explained in the context of GSHS models. In summary, these are their main features:

- **Step 4**: as the model execution for a given vector value $x$ of the stochastic parameters now is another stochastic process, the objective function for the search & partition of Step 6 below has to be an aggregate measure of a set of execution instances. Any aggregate measure is acceptable (e.g. weighted mean, root mean square, etc.) as long as it contributes to finding the regions where the target event occurs and to obtain an acceptable error in the probability estimation of this event.
- **Step 5**: because of the same execution stochasticity, each objective function evaluation may contain a non-null probability of the rare event, instead of a crisp yes/no featured in the basic method of Section II. Thus, a variance reduction method is used within each hyperbox. A particle filtering technique called Interactive Particle System (IPS)[11], [12] is chosen for this purpose. This technique requires the definition of a filtration criterion and corresponding filtration stages, which make feasible the estimation of very low event probabilities.

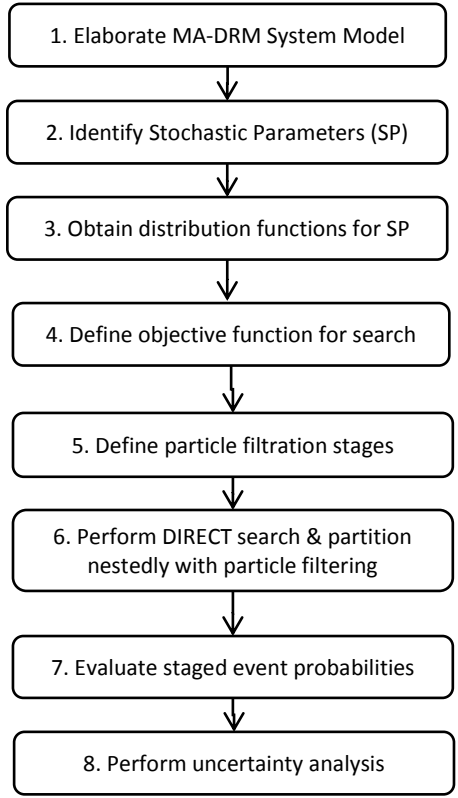

**Figure 11**: The method of rare event probability estimation for models with Stochastic Differential Equations (SDEs).

- **Step 6**: here the DIRECT-based search and partition presented in Section II as Step 5 is combined with the IPS variance reduction technique. Other methods could be used, however the advantage of IPS particle filtering is that no assumptions about equilibrium and absorbing states are needed. The system execution process only has to have the Markov property, which usually holds in physics-based processes outside the sub-atomic domain.
- **Step 7**: the successive filtration stages of IPS allow that the probability not only of the final target event be calculated, but also of the preceding events defined by these filtration stages. This provides a better understanding of the system behavior before the occurrence of the ultimately critical event, which contributes to design improvement insights and implementation of safeguards.

In order to have a better understanding of how these steps happen in practice, in the next section the system model of Section III is enlarged with SDEs of atmospheric turbulence, which significantly changes the model behavior and requires the use of the advanced method features of this section in order to reliably estimate the probability of collision with terrain. These elements as well as estimation results are presented.

## V. CASE STUDY WITH STOCHASTIC DIFFERENTIAL EQUATIONS

This section presents details on the application of the probability estimation method of Figure 11 for a more complex system model with the inclusion of atmospheric turbulence into the example from Section III. This requires a more elaborate algorithm for variance reduction in combination with Monte Carlo, which is the most important contribution of this paper, presented and used in sections V.B through V.D.

### A. Dryden turbulence model

Atmospheric turbulence is a common phenomenon influencing aircraft flight. In the system model developed here, the turbulence model will affect the point-mass model of aircraft dynamics. For this, the Dryden turbulence model was chosen because it is one of the standard models used to evaluate aircraft design [41]. It defines the linear and angular velocity components of air gusts as position-dependent stochastic processes, and is based on the power spectral density of each spatial and angular component. An important characteristic of it is that these power spectral densities are rational, so that they can be implemented as exact filters that take a band-limited white noise input and generate a stochastic process output with filters derived from the Dryden power spectral densities. Thus, if $y_u(s)$ is the power spectral density of the turbulence linear speed component on the dimension $u$, it is modeled according to Dryden as

$$y_u(s) = G_u(s)sW(s) \qquad (2)$$

where $G_u(s)$ is the filter or transfer function, and $W(s)$ is a standard Wiener process (a.k.a. Brownian motion) which, when treated as a *generalized random process* [42], can have its $n$-th order derivatives. The first order derivative $sW(s)$ is white noise.

The filters for the linear speed components $(u, v, w)$, respectively corresponding to the longitudinal, transversal and vertical axis aligned with the aircraft body, in the frequency domain, are

$$G_u(s) = \sigma_u \sqrt{\frac{2L_u}{\pi V}} \frac{1}{1 + \frac{L_u}{V}s}, \qquad (3)$$

$$G_v(s) = \sigma_v \sqrt{\frac{2L_v}{\pi V}} \frac{1 + \frac{2\sqrt{3}L_v}{V}s}{\left(1 + \frac{L_u}{V}s\right)^2}, \qquad (4)$$

$$G_w(s) = \sigma_w \sqrt{\frac{2L_w}{\pi V}} \frac{1 + \frac{2\sqrt{3}L_w}{V}s}{\left(1 + \frac{L_w}{V}s\right)^2}; \qquad (5)$$

where $V$ is the aircraft airspeed, $(\sigma_u, \sigma_v, \sigma_w)$ are turbulence intensity parameters, and $(L_u, L_v, L_w)$ are length parameters.

These parameters are chosen according to the severity scenario to be represented along with the aircraft altitude, and the rules and values are defined in [41]. In the present implementation, fixed values $\sigma_u = \sigma_v = \sigma_w = 7$ ft./sec. were used, corresponding to "light" severity turbulence at the altitude of 4,000 ft., and $L_u = L_v = L_w = 1{,}750$ ft., specified for altitudes above 2,000 ft.

These equations, when transformed to the time domain, result in stochastic differential equations of the form:

$$a_2 \ddot{y}(t) + a_1 \dot{y}(t) + a_0 y(t) + b_2 \ddot{W}(t) + b_1 \dot{W}(t) = 0 \quad (6)$$

where the terms $a_n$ and $b_m$ are defined in terms of the constants used in equations 3-5 and of the aircraft airspeed $V(t)$ and its derivative $\dot{V}(t)$.

The angular components of the turbulence are not represented because the aircraft model here is only point-mass. To illustrate the effect of this turbulence model on the aircraft dynamics, here is the trajectory example of Figure 12.

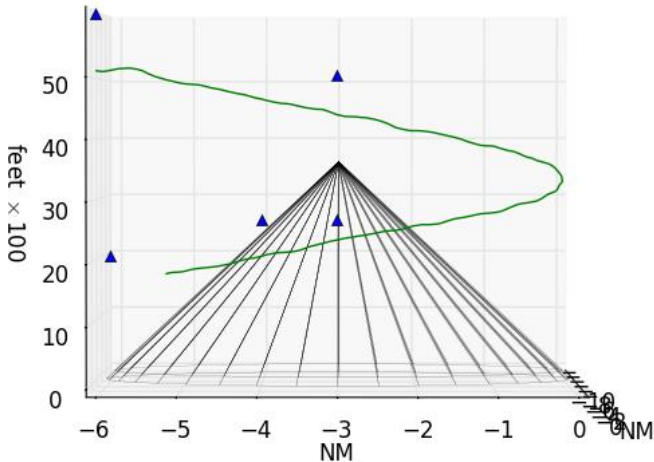

**Figure 12**: Example of turbulent trajectory.

*B. Defining objective function for search and partition*

Regarding Step 4 of Figure 11, the approach chosen here to handle a complex system model is to mask its stochasticity under an aggregate measure for the objective function used in DIRECT. Let $\xi$ denote a stochastic instance of the system, also called a particle, which has a set of state variables, including the vector $\boldsymbol{x}$, which stays fixed during the lifetime of $\xi$. The association of $\xi$ to $\boldsymbol{x}$ is expressed as $\boldsymbol{x} = \boldsymbol{X}(\xi)$. In the model of Section III, the mapping $\boldsymbol{X}$ is invertible, i.e., given a $\boldsymbol{x}$, there is a deterministic and unique $\xi' = \boldsymbol{X}^{-1}(\boldsymbol{x})$ associated to $\boldsymbol{x}$, and $\xi'$ is implicitly used to evaluate $d(\boldsymbol{x})$. But, as $\xi$ is here a stochastic process, there is no longer such uniqueness, $\boldsymbol{X}^{-1}(\boldsymbol{x})$ becoming a random variable. For this reason, it is also no longer possible to use the terrain miss distance $d(\boldsymbol{x})$, as previously defined, because it depends on the instantiation of $\boldsymbol{X}^{-1}(\boldsymbol{x})$.

First, $d(\cdot)$ is redefined to the particle (or trajectory) domain, signifying that it evaluates the terrain miss distance of a concrete trajectory of the system model $\xi$, hence the notation $d(\xi)$. Then, a new function $\bar{d}(\boldsymbol{c}_i) \triangleq E[d(\xi_j)]$ is defined, to be used at each hyperbox $\boldsymbol{B}_i$ centered at $\boldsymbol{c}_i$, for $\xi_j \in \boldsymbol{X}^{-1}(\boldsymbol{c}_i), j = 1, \ldots, s$. This expresses the expectation or mean of evaluations of the distance function $d$ over a number $s$ of system instances associated to $\boldsymbol{c}_i$. This function maintains the convergence of the DIRECT search and partition algorithm, a fact that was observed in practice but can also be mathematically demonstrated.

The new stochastic features of the model include the fact that, when performing the evaluations $d(\xi_j)$, for $\xi_j \in \boldsymbol{X}^{-1}(\boldsymbol{c}_i)$, it may happen that some of the obtained values are higher than $m$ (the distance threshold which defines the target event), and others are equal or lower. The ratio between the number of "hits" $\hat{h}(\boldsymbol{B}_i)$, i.e., the number of instance values equal or lower than $m$, and the total number of instances $s$ at that point, is an estimator of the probability of the system to reach the target region when the input variables assume values in $\boldsymbol{B}_i$, provided $\boldsymbol{B}_i$ is acceptably small. This hit ratio is denoted as $\rho(\boldsymbol{B}_i) = \hat{h}(\boldsymbol{B}_i)/s$ in a crude Monte Carlo definition. Therefore, in the final calculation of the event probability, $\rho(\boldsymbol{B}_i)$ serves as a weighing factor on top of the prior probability of $\boldsymbol{B}_i$.

This means that there is no longer a *border* of the target region, but rather $\rho$-curves and $\rho$-regions, with $\rho \in [0,1]$. The higher the $\rho$, the more frequent the target event being. Thus, the new objective function should promote less subdivision at high plateaus of $\rho$ and concentrate at the slopes that surround the plateaus of $\rho = 1$. Hence, a new definition of $f_o$ was elaborated with a recursive algorithm, in order to obtain this effect. This new objective function received the denomination *Outer* because this feature of concentrating hyperbox subdivision at the outer vicinity of the border of the regions with $\rho = 1$. Its algorithmic definition can be found in the Appendix A. The *Outer* feature contrasts with the $f_o$ defined by Equation 1, which concentrates subdivision at interior points of the target region and will be henceforth denominated by *Inner*.

*C. Definition of particle filtration stages*

The variance reduction technique to be used below, based on particle filtering, needs the definition of a sequence of events which are nested and gradually rarer. In the hypothetical example presented, the final target event is collision with the ground, which happens by definition when $d(\xi) = 0$. However, before this happens, this distance becomes gradually smaller. Starting at $d(\xi) \approx \infty$, the events $d(\xi) \leq m_l$ happen successively with $m_l > 0$ for $l = 1, \ldots, N_F - 1$ and $m_{N_F} = 0$, being $N_F$ the stipulated number of filtration stages. The rationale here is that, given that $d(\xi) \leq m_l$ ($m_0 \approx \infty$) happened, $d(\xi) \leq m_{l+1}$ is not so hard to obtain, thus providing an acceptably large statistical significance. This principle is illustrated by means of Figure 13.

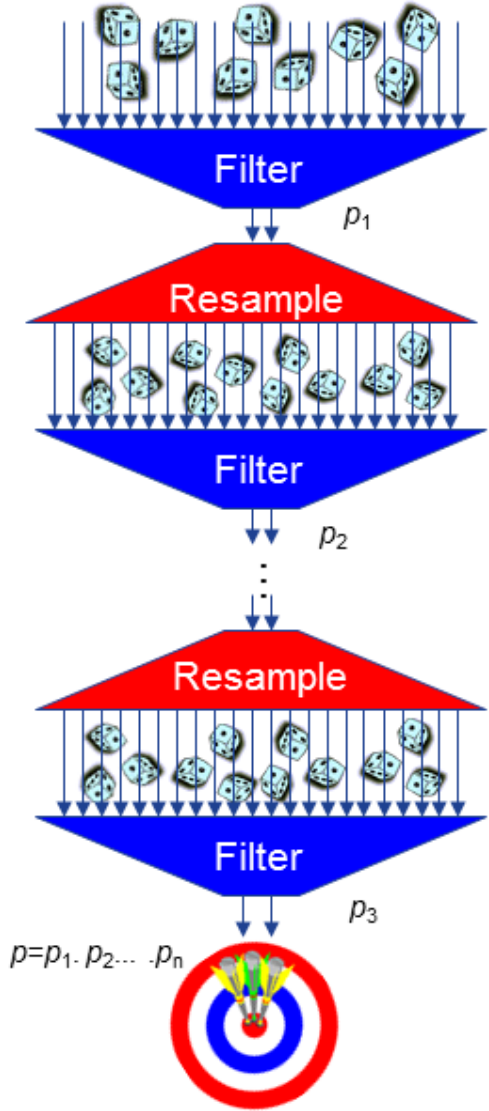

**Figure 13**: Illustration of the principle of particle filtering.

In this figure, $p_l$ represents the probability of a particle surviving the filter of stage $l$, and $p$ the final probability of surviving all stages. Usually, the number of stages needed hangs around the decimal order of magnitude of the probability to be estimated, without the sign. If the probability is approximately 1E–15 ($1 \times 10^{-15}$), the number of stages needed is around 15. The selection of $m_l$ values needs some guessing in the first experiments with a new model, but can be adjusted along preliminary simulation runs. For the hypothetical application example used, the values were set to the ones in Table 1, thus completing Step 5 of the method in Figure 11.

**Table 1**: Distances used to define particle filtering stages.

| Stage index $l$ | Threshold distance $m_l$ (feet) |
|---|---|
| 1 | 1000 |
| 2 | 900 |
| 3 | 800 |
| 4 | 700 |
| 5 | 600 |
| 6 | 450 |
| 7 | 300 |
| 8 | 225 |
| 9 | 150 |
| 10 | 100 |
| 11 | 75 |
| 12 | 50 |
| 13 | 25 |
| 14 | 0 |

*D. Execution of DIRECT search and partition*

Step 6 of Figure 11 differs from its analogous Step 5 from Figure 1 in several ways. As mentioned previously, each evaluation of the objective function $f_o$ chosen for DIRECT contains a rare event process and, because of this rarity, we chose to use the Interactive Particle System (IPS) variation reduction technique [11], [12], in order to determine the event probabilities inside each hyperbox. The central mechanism of IPS is, instead of working with just one region of interest, to use a succession of nested regions as illustrated by means of Figure 13 or, in a formulation of probability theory, a filtration of $\sigma$-algebras of outcomes, among which the innermost corresponds to the final region of interest.

This combined algorithm with DIRECT and IPS is named DIPS, together with the prefix *Outer*, in reference to the newly customized objective function $f_o$. In this algorithm, a separate IPS run is executed inside each hyperbox $\boldsymbol{B}_i$, where the weights $\omega_k^{\langle l \rangle}$ of the particles inside it must be used to account for the prior probability $P_{\boldsymbol{B}_i}$ of the hyperbox (which in turn is based on the density function $g$ of stochastic parameter values).

D.1. Defining a method for determining the confidence interval of the results

At this point, we have all the elements to run the DIPS estimation algorithm. However, because of the high dimensional stochasticity of the system model, we cannot rely on running it a single time. Each time the algorithm runs, a different probability value emerges, and therefore several runs are needed to determine a confidence interval for the target event probability. The histogram of results obtained is highly skewed, due to the high variance of the process modelled, even with the variance reduction technique applied. In this case, using confidence intervals based on normal distributions is not effective, so a different approach is used here, from [43], which is a modified version of the Cox method. Instead of using a standard normal variate $z$ parameter, the $t$ parameter from Student's distribution is used to determine the amplitude of the interval for a given confidence level, in order to better account with small sample sizes. Defining $p = \log(P)$ and $\sigma$ as the sample standard deviation of $p$, the limits of the confidence interval for the true mean of $P$ are given by

$$\exp\left(\bar{p} + \frac{\sigma^2}{2} \pm t\sqrt{\frac{\sigma^2}{N_r} + \frac{\sigma^4}{2(N_r - 1)}}\right) \quad (7)$$

from which the third term is used to define the dispersion measure $\theta$:

$$\theta \triangleq t\sqrt{\frac{\sigma^2}{N_r} + \frac{\sigma^4}{2(N_r - 1)}} \quad (8)$$

D.2. Running the Outer-DIPS estimation algorithm

The dispersion resulting from successive runs of the estimation algorithm is a tradeoff between the time spent to run each algorithm instantiation and the total number of instantiations used in the sample. After an unstructured trial-and-adjust process, the parameters which define the effort in each algorithm instantiation are settled. In these experiments, the number of particles per hyperbox $s$ was set to 1,000 and the total number of hyperboxes to be generated was set to 16,700. We ran the Outer-DIPS algorithm 32 times, in order to gain some benefit from the law of large numbers, and so we obtained the mean and percentile values of Figure 14.

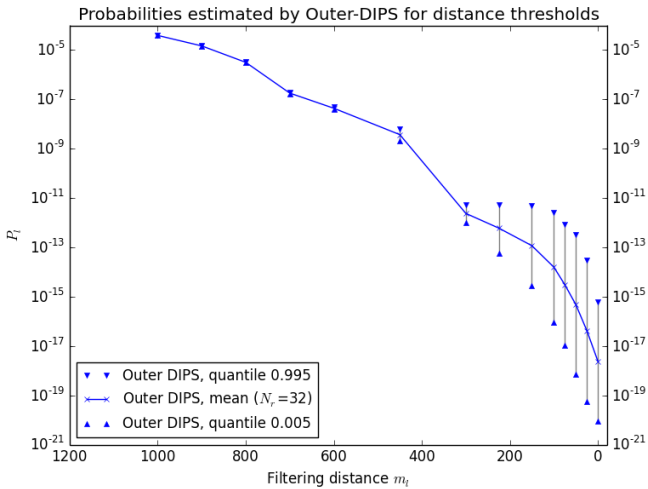

**Figure 14**: Probabilities of hitting filtering distances $m_l$, evaluated by the Outer-DIPS algorithm.

As it can be noted, the confidence intervals, delimited by lower and upper quantiles, are narrow for higher values of $m_l$, but widen as $m_l$ decreases, as seen in the log scale. At $m_l = 0$, the dispersion of the results, calculated according to Equation 8, is 5.05. From a safety analysis viewpoint, the upper quantile is more important, but the fact of having obtained a statistically significant interval for such low probability event gives more credibility to the estimation.

At this point, Steps 6 and 7 of Figure 11 can be considered complete. Further examination of the accuracy and performance of the DIPS algorithm is provided in the Appendix B, and it is worth mentioning that previously existing algorithms failed to obtain the collision event. The last step of the analysis method described in this paper is explained just briefly in the next section.

*E. Considerations about uncertainty analysis*

The confidence interval that we obtained for the probability of the target event is valid for a unique combination of values of the stochastic parameters of the model. Similarly to the basic method of Section II, the knowledge of these values is subject to uncertainties, so the uncertainty analysis of Step 8 of Figure 11 has two main goals: one, reviewing the confidence intervals of the resulting estimate in order to account for these uncertainties; and two, performing parameter sensitivity analysis in order to find the most critical parameters of the system.

The uncertainty analysis capabilities here are almost entirely similar to that explained in Sections II and III, with the differences that here the hit ratio values $\rho(\boldsymbol{B}_i)$ must be maintained for reuse in the probability recalculation, and that, if these values are maintained per filtration stage, the uncertainty analysis may be performed per filtration stage. The computational advantages observed in Step 6 of sections II and III continue to exist at this point.

VI. CONCLUSIONS

The approach proposed in this paper for estimation of the probability of rare events is capable of obtaining statistically significant results for probability values lower than any other in the literature, when considering complex socio-technical systems. Our bibliographic search did not find an example of probability estimation below 1E–10 for such models, while here there are reliable results below 1E–17. For systems without Stochastic Differential Equations (SDEs), the estimation method is highly automatable once the mathematical system model is elaborated. The core automation is provided by the use of a general-purpose search algorithm for continuous variables equipped with an objective function which helps finding the regions where the target event happens. Moreover, the use of the DIRECT algorithm for this aim generates as a by-product a partition of the parameter search space which allows the calculation of very low probabilities, and facilitates both determination of confidence intervals and sensitivity analysis.

If the system model embodies SDEs, it can be classified as Generalized Stochastic Hybrid Systems (GSHS) [32] and has an infinite number of random variables. Therefore it is natural that estimation of rare events for them is considerably more challenging, both in the model elaboration and in the computational use of the model for the probability estimation. In despite of this added difficulty, the case study presented here showed that the method succeeded with an exemplar of such models. In our understanding, this shows that the method proposed is ready to be used in models of similar level of complexity, and models with higher complexity can be processed with increasing levels of computing parallelism.

These aspects are an expression of how our method behaves when we increase the number of random variables which are inputs to the limit state function. In Section III, we provided examples with 2 and 4 random inputs and there observed a sharp increase in the number of function evaluations from one to the other. In Section V, examples with discretized SDEs introduced thousands of input variables, and another large increase in the computing time was observed. Based on these few and distinct examples it is hard to generalize the behavior of our methods with increasing number of dimensions, but the following fragmented facts may help in assessing the performance of our method: when going from 2 to 4 dimensions, the computing time increased about one hundred fold, a large escalation, but lower than theoretically expected, which would be the square of the time. However, when going from 4 to a few thousand dimensions with SDEs, the computational effort was multiplied by also a few thousand times. Thus, for regular variables, our method is highly sensitive to increasing dimensions, however not that so for variables whose existence is due to the time discretization in SDEs.

One of the lessons of this work is that, when dealing with full GSHS dynamics, the model complexity needs to be parsimoniously managed in order to compute results in reasonable times. The examples here used four search variables and the computing time was already significant for a "regular" computer of 12 computing cores. The DIRECT algorithm was reported to have been successfully used with up to thirty search variables [44], however that case was not of a GSHS model. In order to continue this work, it would be interesting to develop more case studies and improve the usability and the flexibility of the simulation software here developed.

ACKNOWLEDGEMENTS

We would like to thank Dr. Henk Blom from NLR and the peer reviewers of this journal for their thoughtful review, their constructive challenges and suggestions, which substantially improved the quality of this work.

# Appendix A: Algorithmic definition of the Outer objective function

In Section *V.B. Defining objective function for search and partition*, the definition of a new objective function $f_o$ for the probability search & partition algorithm was explained in an abstract way. In order to complete that explanation with more implementation details, this section explains its algorithmic definition.

The DIRECT algorithm, in a certain iteration $i$, calculates a centroid point $c_i$ where all hyperboxes from a series of hyperboxes $B_{i,j}$ of decreasing size will be centered, to be generated at iterations $j = i, i+1, \dots, i+n$ (in reality DIRECT may have an arbitrary number of iterations between successive values of $j$; however, for simplicity, these unitary increments can be used when dealing with a single hyperbox centroid). From the definition of $\bar{d}$ given above, the value $\bar{d}(c_i)$ does not change at these iterations, however a new evolving sequence of values $\bar{d}_j(B_{i,j})$ is created in order to account for the fact that the vicinities of these hyperboxes are different. The recursive definition of $\bar{d}_j$ is expressed in Equation 9, which starts by the definition of the *vicinity term* $\bar{v}_j(B_{i,j})$, based on the characteristic of the hyperboxes at the vicinity of the hyperbox $B_{i,j}$. At the first time when $\bar{v}_j(B_{i,j})$ is evaluated ($j = i$), no vicinity information is used, so that only the hit ratio $\rho(B_{i,i})$ and a scaling constant $\lambda$ are used ($\lambda$ was empirically determined as 10 thousand feet); at the next iterations, $\bar{v}_j(B_i) \triangleq \lambda \sum_{z \in Z_{i,j}} [\rho(B_z)]$ is used, where $Z_{i,j}$ is the set of indices pointing to the neighbors of $B_{i,j}$.

$$\bar{v}_j(B_{i,j}) := \begin{cases} \lambda \rho(B_{i,i}) & \text{if } j = i, \\ \lambda \sum_{z \in Z_{i,j}} [\rho(B_z)] & \text{if } j > i. \end{cases} \quad (9)$$

$$\bar{d}_j(B_{i,j}) := \begin{cases} \begin{cases} \bar{d}(B_{i,i}) + \bar{v}_i(B_{i,i}) \text{ for } j = i \\ \bar{d}_{j-1}(B_{i,j}) + \bar{v}_j(B_{i,j}) \text{ for } j > i \end{cases} & \text{if } \rho(B_{i,i}) > 0, \\ \bar{d}(B_{i,i}) & \text{if } \rho(B_{i,i}) = 0. \end{cases}$$

$$f_o(c_i, j) = \bar{d}_j(B_{i,j})$$

The sequence of values $\bar{d}_j(B_{i,j})$ balances the mean distance function $\bar{d}(B_{i,i})$, the vicinity term $\bar{v}_j(B_{i,j})$ and an accumulation of previous values of itself. The original $\bar{d}$ (without subscript) is used to guide the search when $\rho$ is low or next to zero. When $\rho$ is higher, it becomes more significant than $\bar{d}$ and, when $j > i$, it amplifies the vicinity term $\bar{v}_j$. Both the $\bar{v}_j$ and the accumulation from $\bar{d}_{j-1}$ cause a very steep increment in $\bar{d}_j$, which achieves the effect of freezing hyperbox subdivision in the interior points of high $\rho$, and concentrating subdivision at the slopes surrounding these plateaus. Because of this tendency to concentrate outside the plateaus, this objective function receives the denomination *Outer*, in contrast with the $f_o$ defined by Equation 1, which concentrates subdivision at interior points of the target region and is denominated by *Inner*.

# Appendix B: Verification and validation of the method

In order to assess the correctness and check the validity of the methods developed in this work, we compare their results and execution performance with those of pre-existing algorithms, as following. First, we apply the extrapolation method to the same case study of Section III. Then, we compare variations of our methods with the IPS method.

*B.1. Comparison with the extrapolation method*

We ran an experiment using the extrapolation method according to [19]. This method consists of evaluating the same limit state function (or objective function, $f_o$) of our application model in a series of crude Monte Carlo estimation points, each one corresponding to a sample with extremized input variables. Each Monte Carlo point estimate corresponds to an extremization coefficient $k$. Smaller values of $k$ correspond to stronger extremizations of the input variables, which are iso-probabilistically transformed to be normally distributed with mean zero and standard deviation $\sigma = 1/k$. The resulting Monte Carlo hit ratio $\hat{\rho}$ at each point is used to calculate a *reliability index* defined as $\beta = \Phi^{-1}(1-\rho)$, where $\Phi(\cdot)$ represents the cumulative distribution function of the normal probability distribution with $\mu = 0, \sigma = 1$. The limit state function used here for the event estimate is the same as that used in Section III for generating Figure 10. With this setting, the results obtained are shown in Figure 15.

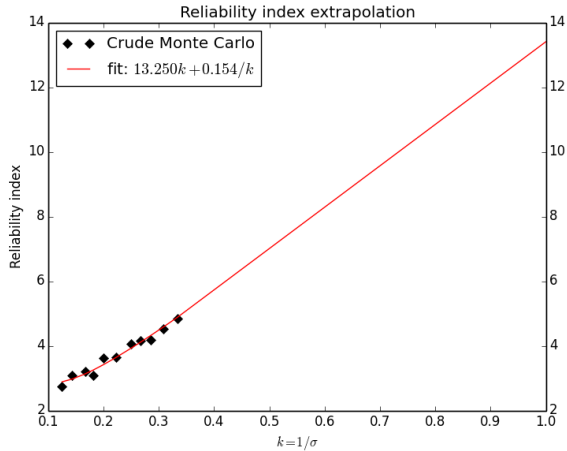

**Figure 15**: Result of the extrapolation method [19] for the case study of Section III.

The number of function evaluations used to generate Figure 15 was approximately 254,000, about the same of that necessary for generating Figure 10 with our method. However, the estimated probability of the rare event is quite different. The extrapolation of Figure 15 corresponding to $k = 1$ is $\beta = 13.4$, and this corresponds to an event probability value of $\rho = 1 - \Phi(13.4) = 1.66 \times 10^{-41}$. Conflicting with this result, our method yields the event probability of $1.66 \times 10^{-21}$. We also evaluated Monte Carlo estimates for $k = 0.4$, but this took several hours and did not change the result significantly. Larger values of $k$ would take even longer.

The reasons for such discrepancy between these methods are not understood to this moment, but because our method is on the conservative side in a risk assessment, we think that it has advantage in this case.

We could not do a comparison with the extrapolation method in the case study of Section IV because our limit state function (based on the simulated aircraft under turbulence) is not easily adaptable to that method. In that limit state function, the thousands of small turbulence disturbance variables act sequentially as input to the aircraft feedback control loop, which can be easily destabilized by the value extremization performed in the extrapolation method. In this case, the majority of collisions with the ground would be consequence of a previous aircraft destabilization, and this event is not the target of the case study, which aims at controlled collisions with terrain. Still, it would be possible to detect destabilization and filter out such cases, or even develop an importance sampler which avoids destabilization. However we think that either case would be a considerable challenge and deemed such tasks as out of scope in this work.

*B.2. The Outer-µ algorithm*

The first estimation algorithm used in the comparison is similar to that of Section III but for a minor modification, which we name here Outer-$\mu$. In this algorithm, a crude Monte Carlo is used in the objective function as well as for the final probability estimation, with threshold distance $m = 0$. The second comparative benchmark is the IPS itself, with fixed filtration criteria; and the third comparative benchmark is IPS with adaptive filtration criteria. The DIPS runs are just the same as those of Figure 14. The determination of confidence quantiles is done accordingly to the method explained in Section V.D.1, with lower quantiles having been omitted form the figure because they were deemed as both irrelevant for a safety analysis and mathematically unreliable, given the extreme skewness found in some of the series. The probability values obtained with these algorithms are shown in Figure 16 and discussed in the following subsections.

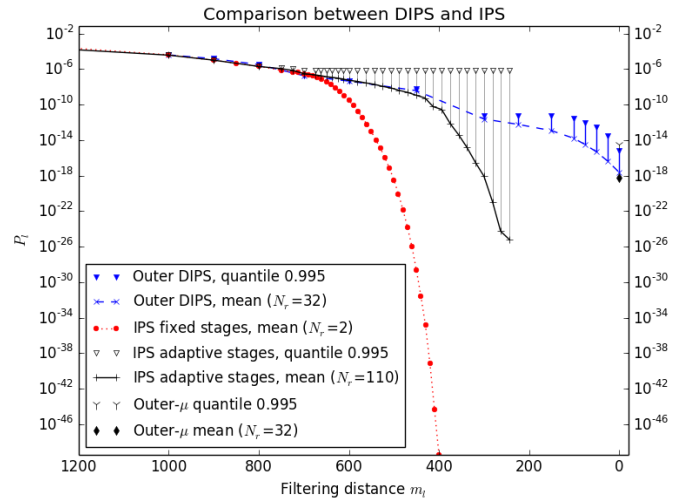

**Figure 16**: Comparison of probabilities of hitting filtering distances $m_l$, according to different algorithms.

As explained above, the Outer-$\mu$ algorithm is the algorithm of Section III, but additionally equipped with a crude Monte Carlo in the objective function and in the final probability estimation, with threshold distance $m = 0$. Setting the algorithm parameters $q = 250,000$ (the total number of hyperboxes) and $s = 100$ (particles per hyperbox), and running it for $N_r = 32$ times, we obtained

**14**

the results shown in **Table 2**. The upper and lower quantiles are according to Equation 7, and the Dispersion indicator according to Equation 8.

**Table 2**: Results from $N_r = 32$ runs of Outer-$\mu$ with $q = 250{,}000$ and $s = 100$.

|  | Sample mean | Dispersion | Quantile 0.005 | Quantile 0.995 |
|---|---|---|---|---|
| $P$: Net hit probability | 6.31E–19 | 6.80 | 3.78E–21 | 3.04E–15 |

Each algorithm run necessary to generate the numbers summarized in this table took an average of 13.5 hours, with little variation. These runs were performed in a multi-core machine, an Intel Xeon X5650, clocked at 2.67 GHz, with 6 physical and 12 logical processing cores, therefore the 32 runs were split in blocks of 12, 12 and 8 simultaneous runs. The dispersion obtained with this algorithm, 6.80 for $m_l = 0$, is a little larger than that obtained by Outer DIPS, which was 5.05. This fact is confirmed visually on Figure 16.

For checking how different the results of Outer DIPS are from those of Outer-$\mu$, a 2-sample t-test was run on the respective samples of probability estimates, for a confidence level of 95% ($t = 1.96$). The logarithm of the sample values was used, because both the sample distributions are skewed to the right, so the test was performed with $\ln(P_l)$ (the natural logarithm), and test statistic $T = 3.02$ was obtained (with a corresponding $p$-value of .0037), which is greater than the critical value of 1.96 and allows to reject the null hypothesis of equal means. The mean probability value estimated with Outer DIPS is 2.41E–18, while that of Outer-$\mu$ is 6.31E–19. This means that DIPS finds the target accident event with more frequency than Outer-$\mu$, a fact that would imply more conservative conclusions in a safety analysis and can be considered a win for DIPS. Adding this to the fact that DIPS has a better precision than Outer-$\mu$ only increases the advantages of DIPS. However, the computational efficiency, involving time and memory, is another criterion that must be analyzed and is discussed in the end of this Appendix.

*B.3. Running IPS with fixed filtration stages*

Because IPS does not partition the probability space, a high number $s$ of particles have to be used to cover the entire search space. In other words, IPS has only one hyperbox which covers the entire search space. Even with $s = 10^6$, most of these IPS trial runs terminated at the first filtration stages, very far from the rare events sought. The first filtration stage had to be set to $m_1 = 2{,}500$ ft., much larger than the $m_1 = 1{,}000$ ft. defined for DIPS, according to Table 1; furthermore, after a tedious trial and error process, we had to considerably increase the number of stages, keeping their threshold values very close, in order to keep the algorithm running with non-null survival rates. The only two valid results obtained with this IPS version are shown in Figure 16, together with the other algorithms in comparison.

Besides these unsatisfactory estimation results, one of these IPS runs took 83 computing hours and another took 48 hours, much longer than any other algorithm in this comparative validation. A 0.79 calibration factor has to be applied to these times, because we had to run this algorithm in another machine, with CPU Xeon E5-1650, with performance a little worse than the previous machine. Even with this large time, we tried increasing the number of particles $s$ to $10^7$. This ate up the whole available computer RAM, with each algorithm instance taking 5 Gigabytes, and we kept 5 of such instances running for 65 days without returning any results, until an accidental shutdown occurred. We could have configured the program to show intermediate results, but the length of the run is by itself a factor which discourages further exploration of this brute force path. The sharp fall of the probability values for distances below than $m_l \approx 600$ ft. just shows that the algorithm cannot generate the rare events sought for lower values of $m_l$.

*B.4. Running IPS with adaptive filtration stages*

The difficulties of running IPS with fixed filtration stages led us to develop an improved version of IPS, where the filtration thresholds are defined in an automated fashion. When running the IPS algorithm for the first time, with fixed filtration criteria, one cannot predict the success rates of the filtration stages. Usually, for each stage a pre-determined number of particles are generated and, if the success rate is high, the surviving particles have to be kept in the memory. Substantial memory occupancy eliminates the possibility of running several algorithm instances in parallel, and resorting to physical storage media would further degrade the performance. On the other hand, if the success rate of a filtration stage is too low, there is little diversity in the next filtration stage and the algorithm tends to terminate prematurely at intermediate stages, as it happened with the IPS with fixed filtration criteria. Besides all this, each filtration stage generates a computational overhead for resampling particles and reloading their contents into the high-speed cache memory and/or CPU registers.

In order to overcome these problems, we devised a more efficient way of performing the IPS particle filtering, without changing its mathematical principle. Instead of sampling a pre-determined number particles at each stage, we keep sampling particles until a fixed number of *surviving* particles is achieved, keeping account of how many particles in total have been sampled, in order to calculate the success rate. Certainly, if the filtration threshold is too stringent, a prohibitively large number of trials would be required, and this is where some adaption method is needed. Adaption was accomplished as follows: establish a maximum number of consecutive trial particles without success; if this number is reached, cancel the stage and set a larger filtration threshold. If this happens in the first stage, this is equivalent to restarting the whole algorithm with a larger initial threshold; if it is an intermediate stage, backtrack to the previous stage and set a closer intermediate threshold. With exception of the first

**15**

stage, this process can be repeated until a minimum difference between stages is reached, after which the sampling may go on indefinitely until the given number of surviving particles is met.

A total of 110 algorithm instances of this algorithm were executed, each one taking on average 13 hours to complete, times the 0.79 calibration factor due to the different computer. The results are shown in Figure 16, in which it can be noticed that their survival rates are much better than those of the fixed-filtration IPS, however still much worse than those of DIPS and Outer-$\mu$. It is possible to notice on that figure that the upper quantiles for this algorithms have a constant value below $m_l = 700$ ft.; this happens because of the following rule that we used: the upper quantile corresponding to a filtering threshold $m_2$ cannot be higher than the quantile corresponding to the threshold $m_1$, if $m_2 \leq m_1$, because in this case the event corresponding to $m_2$ is contained in the event corresponding to $m_1$. The problem is that, despite using the method of Equation 7, which is supposed to work well with skewed distributions, the distributions here are so skewed that the upper quantile according to that method would not make sense (e.g., probability value higher than one).

*B.5. Summary of CPU times*

Below is a summary of computing times of the several methods and dimensionalities presented in this paper, normalized to an Intel Xeon X5650 CPU.

**Table 3**: Summary of computing times and number of function calls (NOFC).

| Method | Dimensions | NOFC | Single core time (secs.) |
|---|---|---|---|
| DIRECT (Sec. III) | 2 vars | 3.5E03 | 4.0E00 |
| DIRECT (Sec. III) | 4 vars | 2.5E05 | 4.1E02 |
| Extrapolation (App. B.1) | 4 vars | 2.5E05 | 4.1E02 |
| DIPS (Sec. V) | 4 vars + SDEs | 5.3E08 | 1.2E06 |
| Outer-$\mu$ (App. B.2) | 4 vars + SDEs | 8.0E08 | 1.6E06 |
| IPS adaptive (App. B.4) | 4 vars + SDEs | 1.1E10 | 4.0E06 |
| IPS fixed (App. B.3) | 4 vars + SDEs | 5.0E07 | 2.2E07 |

The "Dimensions" column lists the number of regular input variables to the limit state function, referred to as "vars", and the indicator "SDEs", which appear when the limit state function is calculated based on Stochastic Differential Equations. In those cases, each 0.1 seconds in the simulated flight time, three Gaussian variables are drawn in order to determine the turbulence incremental disturbances in each spatial dimension at that moment. And, because the total simulated time is variable, the code would need to account the number of dimensions at each function call, which would be considerably more complex with the conditional filtration of IPS, so we did not do that. What can be said is that the simulated flight times, when the particles reach collision, is roughly around 200 seconds, and with $3 \times 10$ variables per second, this implies around 6,000 variables for the function evaluation. However, non-colliding particles fly for about 300 seconds, and degenerated cases might reach up to 1,000 seconds, which is the maximum allowed, although we have not observed such extreme cases.

The varying simulated flight time may account partially for the variation of the average time per function call implied in each case of Table 3. If we divide the computing time by the number of function calls (NOFC), we encounter a lot of variation, and several factors are responsible for it, among which: presence of non-zero wind, non-zero turbulence, IPS filtration setting, dynamic memory allocation, memory page faults, varying simulated time, etc. The most extreme cases are of the IPS with fixed filtration, where the average time per function call is 0.44 s, and the IPS with adaptive filtration, with 3.7E–4 s. We believe that the culprits for such large variation are memory management and the filtration setting, as filtrations occur intra function calls.

It is worth noting that the computing time indicated is "Single core", meaning that that would be the computing time if we ran that experiment in a single CPU core. In practice, all the cases with SDEs run parallelly in each core of a multi-core computer, otherwise the leading times would be much less practical. With exception of the two first cases (DIRECT), all the other methods are highly parallelizable.

We did not include pure crude Monte Carlo simulation in our comparison because a quick estimation shows that it would be infeasible. The most optimistic time for computing a single point of the limit state function, according to the data above, is 3.7E–4 s. The largest collision probability estimated in our experiments, including all methods, is about 1E–17, center of the confidence interval. Thus, the expected time to find a single collision using crude Monte Carlo is the above computing time divided by the above probability is 1.7E13 seconds, or about one million years. Of course one single collision is not enough for statistical significance, and on the other hand this task can be massively parallelized. But such large resource consumption makes us to give up such endeavor.

*B.6. Remarks on the comparative analysis*

Here we digress and discuss about the cases which involve SDEs, which are more relevant for our target applications. From the data shown above, one is drawn to conclude that the Outer DIPS algorithm beats all the other more elementary algorithms, both in time to run and in numeric precision, as measured by statistical dispersion. Despite the fact that one instance of it runs in a time slightly greater (10.3 hours) than the calibrated time of an instance of the adaptive IPS (10.27 hours), we had to run many more instances of the adaptive IPS in order to try to tighten up the dispersion, with little success. Thus, when summing up the time taken by all instances, Outer DIPS won undoubtedly with the shortest time.

The only criterion on which Outer DIPS got a second place is memory usage. While Outer-$\mu$ necessitates less than one hundred Megabyte per run, Outer DIPS requires a few

**16**

hundred, because of its filtering mechanism that stores a large number of particles in the main memory.

LIST OF SYMBOLS

$B_i$: a hyperbox in $\mathbb{R}^n$;

$c_i$: the centroid (center point) of $B_i$, a vector in $\mathbb{R}^n$;

$d(\cdot)$: minimum distance function;

$\epsilon_a$: error of the altimeter measurement;

$\epsilon_h$: altitude error;

$f_o$: objective function of the search & partition algorithm;

$g$: a probability density function in $\mathbb{R}^n$;

$\lambda$: a constant in $\mathbb{R}$, greater than $m$;

$m$: a scalar distance used to define the target event;

$N_F$: number of filtration stages;

$N_r$: number of runs of the estimation algorithm;

$P$: nested net hit probabilities;

$P_{B_i}$: prior probability of hyperbox $B_i$, based on the probability density function $g$;

$q$: total number of hyperboxes;

$\rho(\cdot)$: hit ratio inside a hyperbox;

$s$: number of particles per hyperbox;

$T$: maximum elapsed time in a system instance during simulation;

$t_r$: time elapsed before fault detection by the crew;

$\theta$: dispersion measure for a sample of estimates;

$W(t)$: a Wiener process;

$(w_x, w_y)$: wind horizontal coordinates;

$\omega_k^{(l)}$: weight of a particle $\xi_k$ at IPS stage $l$.

$x$: a parameter value vector in $\mathbb{R}^n$;

$\xi$: a stochastic instance of the system modelled, also called a particle;

$X$: a mapping between the space of system instances and a vector in $\mathbb{R}^n$;